\documentclass[aps,prl,twocolumn,superscriptaddress,nofootinbib,showpacs]{revtex4}
\usepackage{amssymb}
\usepackage{graphicx}
\usepackage{epsf,epsfig}
\usepackage{bm}
\usepackage{bbm}
\usepackage{amsfonts}
\usepackage{amsmath}
\usepackage{graphics}
\usepackage[normalem]{ulem}

\newcommand{\be}{\begin{eqnarray}}
\newcommand{\ee}{\end{eqnarray}}
\newcommand{\ba}{\begin{array}}
\newcommand{\ea}{\end{array}}
\newcommand{\no}{\nonumber}
\newcommand{\tr}{\mbox{tr}}
\newcommand{\Tr}{\mbox{Tr}}

\newcommand{\eps}{\varepsilon}

\newcommand{\bfr}{{\bf r}}

\newcommand{\bfq}{{\bf q}}
\newcommand{\bfs}{{\bf s}}

\newcommand{\bfsigma}{{\bm \sigma}}

\begin{document}

\title{Thermal Transport and Wiedemann-Franz Law in the Disordered Fermi Liquid}
\author{G. Schwiete}
\email{schwiete@uni-mainz.de} \affiliation{Dahlem Center for Complex Quantum Systems and Institut f\"ur Theoretische
Physik, Freie Universit\"at Berlin, 14195 Berlin, Germany
}
\affiliation{Institut f\"ur Physik, Johannes Gutenberg-Universit\"at Mainz, Staudinger Weg 7, 55128 Mainz, Germany}
\author{A. M. Finkel'stein}
\affiliation{Department of Physics and Astronomy, Texas A\&M University, College Station, Texas 77843-4242, USA}
\affiliation{Department of
Condensed Matter Physics, The Weizmann Institute of Science, 76100
Rehovot, Israel}
\affiliation{Institut f\"ur Nanotechnologie, Karlsruhe Institute of Technology, 76021
Karlsruhe, Germany}
\date{\today}

\begin{abstract}
We study thermal transport in the disordered Fermi liquid at low temperatures. Gravitational potentials are used as sources for finding the heat density and its correlation function. For a comprehensive study, we extend the renormalization group (RG) analysis developed for electric transport by including the gravitational potentials into the RG scheme. Our analysis reveals that the Wiedemann-Franz law remains valid even in the presence of quantum corrections caused by the interplay of diffusion modes and the electron electron interaction. In the present scheme this fundamental relation is closely connected with a fixed point in the multi-parametric RG-flow of the gravitational potentials.
\end{abstract}

 \maketitle

\paragraph{Introduction.}
Thermal conductivity $(\kappa)$ measures the ability of a system to conduct heat in response to an applied temperature gradient. In a Fermi liquid, electric and thermal conductivity are tightly linked to each other by the Wiedemann-Franz law (WFL), $\kappa=\mathcal{L}_0 \sigma T$, where $\mathcal{L}_0=\pi^2/3e^2$ is the Lorenz number, $T$ is the temperature and $e$ is the electron charge \cite{Units}. The validity of the WFL in an ordinary Fermi liquid is closely connected with the quasiparticle description \cite{Chester61,Langer62,Kontani03}. At low temperatures, however, transport in disordered conductors is not governed by the rare scattering of quasiparticles on impurities, but rather by diffusive modes and their interactions. The interplay of the slow diffusive modes gives rise to singular quantum corrections to various physical quantities including conductivity, thermal conductivity, specific heat and tunneling density of states \cite{Altshuler85, Finkelstein90}. The question arises about the fate of the WFL in the presence of these strongly interacting collective modes.

Simultaneous measurements of thermal and electric conductivities at low temperatures are frequently used as a means for testing the applicability of the quasiparticle description \cite{Tanatar07,Smith08,Pfau12,Mahajan13,Dong13}. In this context, a violation of the WFL is often interpreted as evidence for physics beyond the Fermi liquid paradigm. Is the reverse statement also true? Do systems with elements of non-Fermi liquid behavior necessarily violate the WFL? Here, we address this question in the context of the singular quantum corrections arising in a disordered Fermi system at low temperatures, $T\ll 1/\tau$, where $1/\tau$ is the elastic scattering rate. The temperature dependences of both thermal conductivity and electric conductivity are strongly affected by these singular corrections. As we will show, thermal and electric transport are nevertheless tightly linked to each other, so that the WFL remains valid even at the lowest temperatures. The system studied in this Rapid Communication provides an example in which elements of non-Fermi liquid behavior are not accompanied by a violation of the WFL.

A systematic treatment of quantum corrections in disordered electron systems has been developed in a renormalization group (RG)  approach on the basis of a field-theoretic description, the nonlinear sigma model (NL$\sigma$M) \cite{Finkelstein83}. The RG analysis with the inclusion of electron electron interactions leads to coupled flow equations for the diffusion constant, the frequency and the interaction constants \cite{Finkelstein83,Castellani84,Finkelstein84,Baranov99}; for a review see \cite{Finkelstein90,Belitz94RMP,DiCastro04,Finkelstein10}. Unfortunately, thermal transport has so far not been considered in the framework of the NL$\sigma$M formalism.
In this work, we study thermal transport in the disordered Fermi liquid \cite{Castellani87, Catelani05} by further developing the NL$\sigma$M approach to the RG analysis.
The main difficulty in the theoretical description of thermal phenomena is that the heat density and heat current operators are more complicated than their analogs for charge transport. For the RG analysis, we introduce time-dependent  ``gravitational potentials" \cite{Luttinger64,Shastry09, Michaeli09} as source fields in the microscopic action. The heat density correlation function can be found by a variation of the action with respect to these source fields. Knowledge of the correlation function then allows one to determine the specific heat and the thermal conductivity. We show how the use of the gravitational potentials can be merged with the NL$\sigma$M formalism. This step requires special care since the gravitational field couples to the disorder term in the action, which, in turn, plays a crucial role for the derivation of the sigma model.

The presence of the gravitational potentials in the NL$\sigma$M complicates the RG analysis. If the Hamiltonian density $h=h_0+h_{int}$ consists of $N$ terms, then the gravitational potentials couple to $N$ different terms in the action. All these terms undergo distinct RG transformations, resulting in the necessity to distinguish the gravitational potentials depending on the part of the Hamiltonian density they couple to. The question arises as to what is the character of the RG-flow, when effectively several potentials are involved. The answer is that the logarithmic corrections originating from energies in the RG interval $(T,1/\tau)$ can be absorbed into the scale-dependent RG charges of the \emph{extended} model, i.e., the model which also includes the gravitational potentials. The calculation reveals that once all corrections are taken into account all gravitational potentials remain unrenormalized: there exists a fixed point in the multi-parametric flow of the gravitational potentials. This implies that after performing renormalizations one may return to the original description of the system but with renormalized Fermi liquid parameters determined by the current scale of the RG procedure. This makes clear why the WFL holds during the course of the RG-procedure.

\paragraph{Keldysh action and the correlation function.}
We start our considerations with the Keldysh partition function $\mathcal{Z}=\int D[\psi^\dagger,\psi] \exp(iS[\psi^\dagger,\psi])$.  The action is first limited to $S=S_k$, where
\be
S_{k}[\psi^\dagger,\psi]=\int_\mathcal{C}dt \int_{\bfr} \left(\psi^\dagger i \partial_t\psi-k[\psi^\dagger,\psi]\right)
\ee
is defined on the Keldysh contour $\mathcal{C}$ \cite{Keldish65,Kamenev11}. Here, $k=h-\mu n$, where $h$ and $n$ are the Hamiltonian density and particle density, $\mu$ is chemical potential, and $\psi=(\psi_\uparrow,\psi_\downarrow)$, $\psi^\dagger=(\psi^*_\uparrow,\psi^*_{\downarrow})$ are vectors of Grassmann fields accounting for the fermionic degrees of freedom with two spin components. A peculiar feature of thermal transport is that the action $S_k$ is determined by the heat density $k$, i.e., precisely by the quantity we study.

We are interested in the retarded heat density correlation function $
\chi_{kk}(x_1,x_2)=-i\theta(t_1-t_2)\langle[\hat{k}(x_1),\hat{k}(x_2)]\rangle_T$, where $x=(\bfr,t)$, $\hat{k}=\hat{h}-\mu\hat{n}$ is the heat density operator and the angular brackets denote thermal averaging. Keldysh's contour technique is very suitable for finding correlation functions of this kind: Introducing fields on the forward ($+$) and backward ($-$) paths of the Keldysh contour, one may define the classical ($cl$) and quantum components ($q$) of the heat density symmetrized over the two branches of the contour, $k_{cl/q}=\frac{1}{2}(k_+\pm k_-)$ \cite{Kamenev11}. Then, the retarded correlation function can be obtained as $\chi_{kk}(x_1,x_2)=-2 i\left\langle k_{cl}(x_1) k_q(x_2)\right\rangle$, where the averaging is with respect to the action $S_k$. After introducing the source term $S_\eta=2\int_x [\eta_2(x)k_{cl}(x)+\eta_1(x)k_q(x)]$ into the action, $S=S_{k}+S_\eta$, one can find $\chi_{kk}$ as
\be
\chi_{kk}(x_1,x_2)&=&\frac{i}{2}\frac{\delta^2 \mathcal{Z}}{\delta \eta_2(x_1)\delta \eta_1(x_2)}.
\ee

The thermal conductivity $\kappa$ can be found from the disorder-averaged correlation function $\langle\chi_{kk}(x_1,x_2)\rangle_{dis}=\chi_{kk}(x_{1}-x_2)$  as \cite{Castellani87}
\be
\kappa =-\frac{1}{T}\lim_{\omega\rightarrow 0}\left(\lim_{q\rightarrow 0}\left[\frac{\omega}{\bfq^2} \mbox{Im}\chi_{kk}(\bfq,\omega)\right]\right).
\ee
This expression is typical for a transport coefficient related to a conserved quantity.

\paragraph{Gravitational potentials and NL$\sigma$M.}
The Hamiltonian density $h=h_0+h_{int}$ is chosen to describe a Fermi liquid in a static disorder potential
\be
h_{0}&=&\frac{1}{2m^*}\sum_{\alpha}\nabla\psi^*_{\alpha}(x)\nabla{\psi}_{\alpha}(x)+u_{dis}(\bfr)n(x),\\
h_{int}&=&\frac{1}{4 }n(x)(\nu^{-1} F_0^\rho) n(x) +\bfs(x) (\nu^{-1} F_0^\sigma) \bfs(x).
\ee
Here, $\nu$ is the density of states per spin, $F_0^{\rho,\sigma}$ are the Fermi-liquid parameters, $m^*$ is the effective mass and $u_{dis}$ is the disorder potential. Further, $\bfs=\frac{1}{2}\sum_{\alpha\beta}\psi^*_\alpha \bfsigma^{\alpha\beta} \psi_\beta$ is the spin density. We anticipate that in the diffusive limit, $T\tau\ll 1$, which we will study here, only the zeroth angular harmonics will be effective.

To proceed further, we perform the Keldysh rotation \cite{Larkin75,Kamenev11} and decouple the interaction terms using a Hubbard-Stratonovich field $\theta_k^l$, where the index $k=1,2$ counts the two Keldysh components ($1,2$ correspond to $cl,q$), and the index $l=0-3$ denotes the density and spin density interaction channels. After this decoupling one can write the action as
\be
{S}&=&\int_x \;\Psi^\dagger \{i\partial_t-[u_{dis}-\mu](1+\hat{\eta})+\hat{\theta}^l\sigma^l\}\Psi \no\\
&&-\int_x\frac{1}{2m^*}\nabla \Psi^\dagger(1+\hat{\eta})\nabla \Psi +\int_x \vec{\theta}^T\frac{\hat{\gamma}_2}{1+\hat{\eta}}f^{-1}\vec{\theta}.\label{eq:S3}
\ee
From now on, $\Psi(x)$ and $\Psi^\dagger(x)$ are fields with two Keldysh components (their spin indices are not shown); the hat symbol indicates matrices in Keldysh space. The matrices $\hat{\theta}$ and $\hat{\eta}$ are defined as $\hat{\eta}=\sum_{k=1,2}\eta_k\hat{\gamma}_k$, $\hat{\theta}^l=\sum_{k=1,2}\theta^l_{k}\hat{\gamma}_k$, where $\hat{\gamma}_1=\hat{\sigma}_0$, $\hat{\gamma}_2=\hat{\sigma}_x$ and $\hat{\sigma}_0$, $\hat{\sigma}_x$ are Pauli matrices in Keldysh space. The Pauli matrices $\sigma_l$ in Eq.~\eqref{eq:S3} act in spin space. The matrix $f=\mbox{diag}(F_0^\rho,F_0^\sigma,F_0^\sigma,F_0^\sigma)/2\nu$ distinguishes the different interaction channels.

The disadvantage of the representation in Eq.~\eqref{eq:S3} is that the gravitational potentials couple to the disorder potential $u_{dis}$, thereby complicating the derivation of the NL$\sigma$M. In the following manipulations we exploit the structural similarity between the source term and the $k$-term in the action. We use this fact to devise a transformation that releases the disorder term from the explicit dependence on the gravitational fields \cite{CommentTrafo}. After that, the $\sigma$-model can be derived following the conventional scheme. The mentioned transformation reads as $\psi\rightarrow \sqrt{\hat{\lambda}}\psi$, $\bar{\psi}\rightarrow \bar{\psi} \sqrt{\hat{\lambda}}$, where $ \hat{\lambda}=1/(1+\hat{\eta})$. (The arising Jacobian is featureless; its only function is to remove disconnected contributions proportional to the heat density itself.) Since the gravitational potentials can be considered as arbitrarily slow, a term proportional to $(\nabla\hat{\eta})^2$ emerging from this transformation may be ignored. As a result, the gravitational potentials are removed from $h_{0}-\mu n$ at the expense of introducing source fields into the time-derivative term and a change in the structure of the interaction part
\be
S&=&\frac{1}{2}\int_x {\Psi}^\dagger(i\hat{\lambda}\overrightarrow{\partial}_t-i\overleftarrow{\partial}_t\hat{\lambda})\Psi
-\int_x \;\Psi^\dagger (u_{dis}-\mu-\hat{\lambda}\hat{\theta}^l\sigma^l)\Psi\no\\
&&-\int_x\frac{1}{2m^*}\nabla \Psi^\dagger\nabla\Psi+\int_x \vec{\theta}^T(\hat{\gamma}_2\hat{\lambda})f^{-1}\vec{\theta}.
\ee
Most importantly, the disorder part of the action does not contain the gravitational potentials anymore. From here on, the NL$\sigma$M can be derived along the standard lines \cite{Wegner79,Efetov80,Finkelstein90, Kamenev99, Chamon99,Schwiete14}; it may be written as $S=S_{dm}+S_{\eta\eta}$, where
\be
S_{dm}&=&\frac{\pi\nu i}{4}\Tr[D(\nabla\hat{Q})^2+2iz\{\hat{\eps},\hat{\lambda}\} \underline{\delta \hat{Q}}],\no\\
&&+\frac{i}{2}(\pi\nu)^2 \langle \Tr[\hat{\lambda}\hat{\theta}^l \sigma^l \underline{\delta\hat{Q}}]\Tr[\hat{\theta}^k\sigma^k\underline{\delta \hat{Q}}]\rangle.\quad
\label{eq:Sd}
\ee
Here, $\hat{Q}$, $\delta\hat{Q}$, $\hat{\lambda}$ and $\hat{\theta}$ are matrices in Keldysh and spin space as well as in the frequency domain. In particular, $(\hat{\lambda}_{\bfr})_{\eps\eps'}=\hat{\lambda}_{\bfr,\eps-\eps'}$ and the same for $\hat{\theta}$, while $\hat{Q}_{\eps\eps'}$ generally depends on both frequency arguments. The structure of $\delta\hat{Q}$ will be specified further below. $\Tr$ covers all degrees of freedom including spin as well as integration over coordinates. The brackets symbolize the contractions $\langle \theta^0_{k,\bfr,\omega}\theta^{0}_{l,\bfr',-\omega'}\rangle=\frac{i}{2\nu}(\Gamma_{\rho}/2)\gamma_2^{kl}\delta_{\bfr-\bfr'}2\pi\delta_{\omega-\omega'}$ and for spin degrees of freedom $\langle \theta^\alpha_{k,\bfr,\omega}\theta^{\beta}_{l,\bfr',-\omega'}\rangle=\frac{i}{2\nu}
(\Gamma_{\sigma}/2)\gamma_2^{kl}\delta_{\bfr-\bfr'}2\pi\delta_{\omega-\omega'}\delta_{\alpha\beta}$, where $\Gamma_{\rho}=F_0^\rho/(1+F_0^\rho)$, $\Gamma_{\sigma}=F_0^\sigma/(1+F_0^\sigma)$.
Finally, note the parameter $z$ in the frequency term anticipating its renormalization in the presence of the electron-electron interaction \cite{Finkelstein83}; the initial value is $z=1$. The charge $z$ plays a central role for thermal transport \cite{Castellani87}, and at the metal-insulator transition \cite{Finkelstein83a,Finkelstein84a, Punnoose05}. The abandoned term $S_{\eta\eta}$ is quadratic in the source fields, and accounts for the contribution of fermions to the static part $\chi_{kk}^{st,0}$. It is disconnected from the diffusion modes, which are described by $S_{dm}$.

The matrix $\hat{Q}$ can be parametrized as $\hat{Q}=\hat{U}\hat{\sigma}_3 \hat{\bar{U}}$, where $\hat{U}\hat{\bar{U}}=1$; the deviations $\delta \hat{Q}=\hat{Q}-\hat{\sigma}_3$ describe diffusive degrees with energies $\lesssim1/\tau$. For $\underline{\delta \hat{Q}}(\eps\eps')=u_{\eps}\delta\hat{Q}_{\eps\eps'}u_{\eps'}$ the temperature of electrons enters through the distribution function encoded in $\hat{u}$:
\be
\hat{u}_\eps=\left(\ba{cc} 1&\mathcal {F}_\eps\\0&-1\ea\right),\quad \mathcal {F}_\eps=\tanh\left(\frac{\eps}{2T}\right).
\ee
The (retarded) diffusive propagation is described by $\mathcal{D}(\bfq,\omega)=1/({D\bfq^2-iz\omega})$, the so-called diffuson.
\paragraph{Specific heat.}
In order to illustrate the use of the gravitational potentials, we start our discussion with the calculation of the specific heat $c$. It comprises a trivial electronic part $c^0$ and a contribution of diffusion modes $c^{dm}$, which we are interested in. Note that the diffusion modes give rise to the heat density $k^{dm}_{\eta_1}(x_1)=(i/2)\delta \mathcal{Z}^{dm}/\delta \eta_2(x_1)|_{\eta_2=0}$, where $\mathcal{Z}^{dm}$ is determined by $S_{dm}$ of Eq.~\eqref{eq:Sd}. To find $k^{dm}_{\eta_1}$, we have to expand $\hat{\lambda}=
1-\eta_1-\eta_2(1-2\eta_1)\hat{\gamma_2}$ in Eq.~\eqref{eq:Sd}. Taking the derivative with respect to $\eta_2$ results in two terms determining the heat density of the diffusion modes; one term originating from the frequency part and the other one from the interaction part of the action. The specific heat can be found directly by differentiating the heat density with respect to temperature \emph{in the absence} of the classical gravitational potential $\eta_1$, i.e., from $k^{dm}_{\eta=0}$. Calculating the two terms for $k^{dm}_{\eta=0}$ in the Gaussian approximation, we find
\be
k^{dm}_{\eta=0}&=&-\frac{1}{2}\int_{\bfq,\omega} z \omega\mathcal{B}_\omega\left(\mathcal{D}-\mathcal{D}_1+3(\mathcal{D}-\mathcal{D}_2)\right)\no\\
&&-\frac{1}{2}\int_{\bfq,\omega}\;\omega\mathcal{B}_{\omega}\left(\Gamma_\rho\mathcal{D}_1+
3\Gamma_\sigma\mathcal{D}_2\right)\label{eq:keta}.
\ee
Here, we introduced propagators for diffusion in the singlet and triplet spin channel, $\mathcal{D}_{1,2}=1/(D\bfq^2-iz_{1,2}\omega)$, where $z_1=z-\Gamma_\rho$, $z_2=z-\Gamma_\sigma$; $\mathcal{B}_\omega=\cot(\omega/2T)$ is the bosonic distribution function. Further manipulations allow us to present the heat density in the form
$k^{dm}_{\eta=0}=\frac{1}{2}\int_{\bfq,\omega}\omega\mathcal{B}_\omega D\bfq^2\left[z_1\mathcal{D}_1\overline{\mathcal{D}}_1+
3z_2\mathcal{D}_2\overline{\mathcal{D}}_2-4z\mathcal{D}\overline{\mathcal{D}}\right]$. According to this formula, the heat density of diffusons is determined by the energy weighted with the distribution function and multiplied by the spectral function of the diffusion modes. Differentiation with respect to temperature gives $c^{dm}=\partial_T k^{dm}_{\eta=0}$. The integrals obtained after differentiation are logarithmic and depend on parameters which are themselves determined by the RG flow. The analysis of such quantities has to be performed in the framework of the RG. The contribution of fermions stays inert in the present discussion. Analysis of the fermionic and the diffusion mode parts of the specific heat leads to the conclusion \cite{Castellani86} that in the disordered Fermi liquid as a result of renormalizations $c=zc_{FL}$, where $c_{FL}=2\pi^2\nu T/3$.

Generally, we are interested in the correlation function $\chi_{kk}$ which can be decomposed into a static and a dynamical part, $\chi_{kk}=\chi_{kk}^{st}+\chi_{kk}^{dyn}$. As we shall see below, the static part is directly related to the specific heat as $\chi_{kk}^{st}=-cT$. For finding the thermal conductivity $\kappa$, in turn, it will be sufficient to know $\mbox{Im}\chi_{kk}(\bfq,\omega)=\mbox{Im}\chi^{dyn}_{kk}(\bfq,\omega)$. Our study of $\chi_{kk}(\bfq,\omega)$ will be based on an RG-treatment in the presence of the gravitational potentials, keeping in mind their dependence on $\bfq$ and $\omega$.

\paragraph{RG analysis in the presence of the gravitational potentials.}
For the discussion of the dynamical part of the correlation function it is sufficient to expand $\hat{\lambda}\approx 1-\hat{\eta}$ in the action. We study here the renormalization of the sources generated by $\eta_1$. It will be preferable to use the interaction amplitudes in the form $\frac{1}{2}(\Gamma _{\rho }\delta _{\alpha \delta }\delta _{\beta \gamma }+\Gamma _{\sigma }\bfsigma_{\alpha \delta }\bfsigma_{\beta \gamma })=\Gamma _{1}\delta _{\alpha \delta }\delta _{\beta \gamma }-\Gamma _{2}\delta _{\alpha \gamma }\delta _{\beta \delta }$, where $\Gamma _{1}=\frac{1}{2}(\Gamma _{\rho }-\Gamma _{\sigma })$ and $\Gamma_2=-\Gamma_{\sigma}$. To this end, one should consider the following action
\be
S_{\zeta}=\frac{\pi\nu i}{4} \Tr[D(1+\underline{\hat{\zeta}_D})
(\nabla \hat{Q})^2+2iz \{\hat{\eps},1+\underline{\hat{\zeta}_z} \}\delta\hat{Q}]\no\\
+\frac{i}{2}(\pi\nu)^2 \sum_{n=1}^2\;\langle \Tr[(1+\underline{\hat{\zeta}_{\Gamma_n}})\underline{\hat{\phi}_n}\delta\hat{Q}]\Tr[\underline{\hat{\phi}_n}\delta\hat{Q}]\rangle,\quad
\ee
where $\underline{\hat{\zeta}_{X}}(\bfr,\eps+\omega,\eps)=\hat{u}_{\eps+\omega}\hat{\gamma}_1\hat{u}_{\eps} \zeta_{X}(\bfr,\omega)$ for $X\in\{D,z,\Gamma_1,\Gamma_2\}$. In the following we shall also use notations $\zeta_i$ and $X_i$ with $i=1...4$. The contractions for the fields $\phi_n$, $n=1,2$  generate the proper interaction terms with $\Gamma_1$ and $\Gamma_2$. The initial conditions are obtained from a comparison with Eq.~\eqref{eq:Sd}, $
    \quad \zeta_z=\zeta_{\Gamma_1}=\zeta_{\Gamma_2}=-\eta_{1},\quad \zeta_D=0$. The field $\zeta_D$ was introduced to account for the possibility that the sources migrate to the kinetic term during the RG procedure.

The general structure of the RG-corrections is determined by the number of independent integrations over momenta. Each integration leads to an additional power in the inverse dimensionless conductance, which is the small parameter of the RG expansion. At a given order of the RG expansion, the dependence on the interaction amplitudes can be accounted for to all orders once the described dressing of the interaction amplitudes is included \cite{Finkelstein83,Finkelstein10}. Therefore, in order to remain within a given order, it is sufficient to extract the $\zeta_X$-terms from the established RG diagrams. The procedure is relatively simple if one deals with potentials $\underline{\zeta_X}(\eps,\eps')$ carrying two fast frequency arguments, since then it is sufficient for the RG to approximate $\underline{\hat{\zeta}_X}(\eps,\eps')\approx \zeta_X(\eps-\eps')$ and at the same time matrices $U$ or $\bar{U}$ with arguments $\eps$ and $\eps'$ may be set equal to $1$. As a result, the extraction of potentials $\zeta_z$ and $\zeta_D$ is essentially realized by a differentiation of the diffusion propagators as $D\partial_D\mathcal{D}$ or $z\partial_z\mathcal{D}$. In a similar way, the extraction of $\zeta_{\Gamma_n}$ may be implemented by a differentiation with respect to $\Gamma_n$.

Unfortunately, if the frequency arguments of $\underline{\zeta_X}(\eps,\eps')$ are slow one has to perform a tedious calculation complicated by the fact that in products of the form
$\hat{\bar{U}}\underline{\zeta_X}\hat{U}$ the matrices $\hat{\bar{U}}$ and $\hat{U}$ remain intact: $\hat{\bar{U}}\underline{\zeta_X} \hat{U}\neq \underline{\zeta_X}$. Still, the above remarks allow one to understand why the final result of the RG-analysis acquires a very compact form:
\be
\Delta (X_{i_{0}}\zeta_{i_{0}})=\sum_{j=1}^4 \zeta_j X_j\frac{\partial}{\partial X_j} (\Delta X_{i_{0}}),\label{eq:compactr}
\ee
where $\Delta X$ symbolizes a logarithmic correction to $X$. The result, which holds for all $X\in\{D,z,\Gamma_1,\Gamma_2\}$, bears a certain resemblance with the multiplicative RG \cite{Bogoliubov59}.

One can show, using the known RG-equations for the charges $X_i$, that the initial values for the sources \emph{do not change} as a result of renormalization. Indeed, the RG-equations in the absence of sources have a rigid structure dictated by the NL$\sigma$M:
\be
{dG}/{d\xi}=\beta\left[G;w_2,w_1\right],
\quad {dY_i}/{d\xi}=z\beta_i\left[G;w_2,w_1\right].\label{eq:RGstr}
\ee
where $Y_i\in\{z,\Gamma_1,\Gamma_2\}$, $G=4\pi\nu D$ and $w_i=\Gamma_i/z$. Then, it follows immediately from Eqs.~\eqref{eq:compactr} and \eqref{eq:RGstr} that the parameters $\zeta_X$ do not flow, $\it{provided}$ that $\zeta_D=0$ holds initially and all remaining $\zeta_Y$ are equal. Note the important fact that $\zeta_D$ cannot be generated by other sources if they are equal.
Thus, we obtained a fixed point in the multi-parametric RG-flow, which is a rather non-trivial result for a multi-parametric flow.

\paragraph{Static part of $\chi_{kk}$:}
In analogy to $c$, we decompose the static correlation function $\chi_{kk}^{st}$ into two parts: $\chi_{kk}^{st}=\chi^{st,0}_{kk}+\chi^{st,dm}_{kk}$, where $\chi_{kk}^{st,0}$ is the trivial electronic part, while $\chi_{kk}^{st,dm}(x_1,x_2)={\delta k^{dm}_{\eta_{1}}(x_1)}/{\delta \eta_1(x_2)}|_{\eta_{1}=0}$ originates from the diffusion modes.
We may use $k^{dm}_{\eta=0}$ as a starting point for the calculation of $\chi_{kk}^{st,dm}$. The terms originating from the expansion of $\hat{\lambda}$ up to $2\eta_1\eta_2\hat{\gamma_2}$, obviously, yield $-2k^{dm}_{\eta=0}$. The remaining terms can be obtained according to the following reasoning. One needs to restore the dependence on the field $\eta_1$ in $k^{dm}$ and extract $\eta_1$ from any part of the diffusons contributing to $k^{dm}_{\eta_1}$. Since the differentiation with respect to $\eta_1$ can be written as a differentiation with respect to the charges $z$ and $\Gamma_i$, one can apply the operator $O_{\mathcal{D}}^\eta=-z\partial_z-\Gamma_\rho\partial_{\Gamma_\rho}-\Gamma_\sigma\partial_{\Gamma_\sigma}$ which acts only on the diffusons $\mathcal{D}$, $\mathcal{D}_{1,2}$. Here we exploit the previously mentioned fact that during the course of the RG procedure, the parameters $Y_i$ follow their ``host" amplitudes and that $\zeta_D=0$. The final result can be written as
\be
\chi^{st,dm}_{kk}=(O_{\mathcal{D}}^\eta-2)k^{dm}_{\eta=0}.
\ee
Using the fact that in the diffusons $z$ and $\Gamma_i$ stand together with frequency $\omega$, one may replace $O_{\mathcal{D}}^\eta-2$ by $\omega(\omega\partial_\omega+2)$, where the differentiation is still restricted to the diffusons.
Next, we make use of the relations $\omega(\omega\partial_\omega+2)f(\omega)=\partial_\omega(\omega^2 f(\omega))$
and $\omega\partial_\omega f(\omega/2T)=-T\partial_Tf(\omega/2T)$
in order to find that $\chi_{kk}^{st,d}=-T\partial_T k^d_{\eta=0}$.
It means that together with the contribution from electrons we indeed have $\chi_{kk}^{st}=-Tc$.

\paragraph{Heat conductivity:} After all renormalizations, the dynamical part $\chi_{kk}^{dyn}$ can be found by averaging the product of the $\eta_{1}$-, $\eta_2$-frequency terms in the ladder approximation. This last averaging generates a diffuson $\mathcal{D}(\bfq,\omega)$,
\be
&&\chi_{kk}^{dyn}=-\frac{i}{8}(\pi\nu)^2 z_{\eta_2}z_{\eta_1}\left\langle \delta _{\eta_2}\tr[\{\hat{\eps},\underline{\eta_2\hat{\gamma}_2}\}\delta\hat{Q}]\right.\label{eq:chidyn}\\
&&\qquad\left.\times \delta_{\eta_1}\tr[\{\hat{\eps},\underline{\eta_1}\}\delta\hat{Q}]\right\rangle=-c_{FL}Tz_{\eta_2} \frac{iz\omega}{D\bfq^2-iz\omega}.\no
\ee
In the last line we used that, as we have shown, the renormalization of the $\eta_1$ vertex is given by $z_{\eta_1}=z$.
The calculation of $z_{\eta_2}$ is beyond the scope of this Rapid Communication. Instead we rely on the fact that for a conserved quantity the sum of the static and dynamical parts of the correlation function vanishes in the limit $q\rightarrow 0$. As we have demonstrated above, $\chi_{kk}^{st}=-Tzc_{FL}$. Then, we come to the known structure of the correlation function \cite{Castellani87}:
\be
\chi_{kk}(\bfq,\omega)=-Tc\frac{D_k\bfq^2}{D_k\bfq^2-i\omega},
\ee
where $D_k=D/z$ is the heat diffusion coefficient. It follows for the thermal conductivity that $\kappa=cD_k=c_{FL}D$.  In combination with the RG results for the conductivity of the disordered Fermi liquid, $\sigma=2e^2\nu D$, this yields the WFL: $\kappa/\sigma=\pi^2 T/3e^2$.

The use of the ladder approximation in Eq.~\eqref{eq:chidyn} amounts to a restriction to collisionless kinetics. While the full NL$\sigma$M of Eq.~\eqref{eq:Sd}, in fact, incorporates collisions, it can been checked that the inclusion of collisions does not lead to additional logarithmic corrections in the model of fermions with a short range interaction.

\paragraph{Conclusion:} By incorporating Luttinger's gravitational potentials into the NL$\sigma$M formalism, we developed a consistent theory of thermal transport for the disordered Fermi liquid \cite{CommentLongRange}. The obtained results imply that in the studied system the WFL remains valid despite the multitude of singular quantum corrections arising at low temperatures. This example clearly demonstrates that the observation of the WFL by itself does not guarantee the applicability of the conventional Fermi liquid description.

\section*{Acknowledgments}
The authors gratefully acknowledge the support by the Alexander von Humboldt Foundation. A.~F.~thanks the members of the Institut f\"ur Theorie der Kondensierten Materie at KIT for their kind hospitality. A.~F. is supported by the National Science Foundation Grant No. NSF-DMR-1006752.


\begin{thebibliography}{38}
\bibitem[{Com({\natexlab{a}})}]{CommentTrafo}
\bibinfo{note}{A somewhat related transformation was used in
  Ref.~\protect{\cite{Michaeli09}}.}
\end{thebibliography}
\end{document}